\documentclass[12pt,twoside,a4paper,twocolumn]{article}
\usepackage{cmap}
\usepackage[T2A]{fontenc}
\usepackage[utf8]{inputenc}
\usepackage[russian,english]{babel} 
\newcommand{\ru}{\selectlanguage{russian}}
\newcommand{\en}{\selectlanguage{english}}
\en
\usepackage[unicode,hidelinks,breaklinks=true]{hyperref}
\usepackage[margin=0.75in,top=1in]{geometry}
\usepackage{textcase}
\usepackage[ruled,lined,linesnumbered,longend]{algorithm2e} 
\usepackage{algpseudocode}
\usepackage{amsfonts}
\usepackage{amssymb}
\usepackage{graphicx}
\usepackage{textcomp}
\usepackage{todonotes}
\usepackage{xcolor}
\usepackage{xifthen}
\usepackage{icomma}
\usepackage[author={dluciv}]{pdfcomment}

\usepackage{fancyhdr}
\usepackage{amsthm}
\usepackage{amsmath}

\newtheorem{theorem}{Theorem}
\newtheorem{definition}{Definition}
\newtheorem{remark}{Note}

\newcommand\blfootnote[1]{%
  \begingroup
  \renewcommand\thefootnote{}\footnote{#1}%
  \addtocounter{footnote}{-1}%
  \endgroup
}

\author{\textbf{D.V.~Luciv${}^1$, D.V.~Koznov${}^1$, G.A.~Chernishev${}^1$, A.N.~Terekhov${}^1$}
\\
${}^1$~\emph{Saint Petersburg State University}\\
{\slshape 199034} {\itshape St. Petersburg, Universitetskaya Emb.,} {\slshape 7--9}\\
\emph{E-mail:~\{d.lutsiv, d.koznov, g.chernyshev, a.terekhov\}@spbu.ru}
}

\title{Detecting Near Duplicates in~Software Documentation}

\renewenvironment{abstract}
 {\small
  \begin{center}
  \bfseries \abstractname\vspace{-.5em}\vspace{0pt}
  \end{center}
  \list{}{%
    \setlength{\leftmargin}{10mm}
    \setlength{\rightmargin}{\leftmargin}%
  }%
  \item\relax}
 {\endlist}

\begin{document}
\twocolumn[{%
\maketitle
\thispagestyle{empty}
\begin{abstract}
Contemporary software documentation is as complicated as the software itself. During its lifecycle, the documentation accumulates a lot of ``near duplicate'' fragments, i.e. chunks of text that were copied from a single source and were later modified in different ways. Such near duplicates decrease documentation quality and thus hamper its further utilization. At the same time, they are hard to detect manually due to their fuzzy nature.
In this paper we give a formal definition of near duplicates and present an algorithm for their detection in software documents. This algorithm is based on the exact software clone detection approach: the software clone detection tool Clone Miner was adapted to detect exact duplicates in documents. Then, our algorithm uses these exact duplicates to construct near ones.
We evaluate the proposed algorithm using the documentation of 19 open source and commercial projects. Our evaluation is very comprehensive~--- it covers various documentation types: design and requirement specifications, programming guides and API documentation, user manuals. Overall, the evaluation shows that all kinds of software documentation contain a significant number of both exact and near duplicates. Next, we report on the performed manual analysis of the detected near duplicates for the Linux Kernel Documentation. We present both quantative and qualitative results of this analysis, demonstrate algorithm strengths and weaknesses, and discuss the benefits of duplicate management in software documents.

\end{abstract}
}]

{}\blfootnote{This work is partially supported by RFBR grant No 16-01-00304.}

\section{Introduction}

Every year software is becoming increasingly more complex and extensive, and so does software documentation. During the software life cycle documentation tends to accumulate a lot of duplicates due to the copy and paste pattern. At first, some text fragment is copied several times, then each copy is modified, possibly in its own way. Thus, different copies of initially similar fragments become “near duplicates”. Depending on the document type~\cite{Parnas2011}, duplicates can be either desired or not, but in any case duplicates increase documentation complexity and thus, maintenance and authoring costs~\cite{juergens2010}.

Textual duplicates in software documentation, both exact and near ones, are extensively studied~\cite{juergens2010,nosal2016,koznov2008,romanovsky2008}. However, there are no methods for detection of near duplicates, only for exact ones and mainly using software clone detection techniques ~\cite{juergens2010,nosal2016,wingkvist2010}. At the same time Juergens et al. ~\cite{juergens2010} indicates the importance of near duplicates and recommends to ``pay particular attention to subtle differences in duplicated text''. However, there are studies which addressed near duplicates, for example in~\cite{nosal2014} the authors developed a duplicate specification tool for JavaDoc documentation. This tool allows user to specify near duplicates and manipulate them. Their approach was based on an informal definition of near duplicates and the problem of duplicate detection was not addressed. 
In our previous studies~\cite{psi15,prog16} we presented a near duplicate detection approach. Its core idea is to uncover near duplicates and then to apply the reuse techniques described in our earlier studies~\cite{koznov2008,romanovsky2008}. Clone detection tool Clone Miner~\cite{basit2009} was adapted for detection of exact duplicates in documents, then near duplicates were extracted as combinations of exact duplicates. However, only near duplicates with one variation point  were considered. In other words, the approach can detect only near duplicates that consist of two exact duplicates with a single chunk of variable text between them: $exact_1 ~ variable_1 ~ exact_2$.

In this paper we give the formal definition of near duplicates with an arbitrary number of variation points, exhibiting the following pattern: $exact_1 ~ variable_1 ~ exact_2 ~ variable_2 ~ exact_3 \dots\allowbreak variable_{n-1} ~ exact_n$. Our definition is the formalized version of the definition given in the reference ~\cite{Bassett1996}. We also present a generalization of the algorithm described in ~\cite{psi15,prog16}. The algorithm is implemented in the Documentation Refactoring Toolkit~\cite{DRT}, which is a part of the DocLine project~\cite{koznov2008}. In this paper, an evaluation of the proposed algorithm is also presented. The documentation of 19 open source and commercial projects is used. The results of the detailed manual analysis of the detected near duplicates for the Linux Kernel Documentation~\cite{linuxkd2013} are reported.

The paper is structured as follows. \textbf{Section~1} provides a survey of duplicate management for software documentation and gives a brief overview of near duplicate detection methods in information retrieval, theoretical computer science, and software clone detection areas. \textbf{Section~2} describes the context of this study by examining our previous work to underline the contribution of the current paper. In \textbf{Section~3}~the formal definition of the near duplicate is given and the near duplicate detection algorithm is presented. Also, the algorithm correctness theorem is formulated. Finally, \textbf{Section~4} presents evaluation results.
\section{Related work}\label{sec:relwork}

Let us consider how near duplicates are employed in documentation-oriented software engineering research.
Horie et al.~\cite{horie2010} consider the problem of text fragment duplicates in Java API documentation. The authors introduce a notion of crosscutting concern, which is essentially a textual duplicate appearing in documentation. The authors present a tool named CommentWeaver, which provides several mechanisms for modularization of the API documentation. It is implemented as an extention of Javadoc tool, and provides new tags for controlling reusable text fragments. However, near duplicates are not considered, facilities for duplicate detection are not provided. 

Nos\'{a}l and Porub\"{a}n~\cite{nosal2014} extend the approach from~\cite{horie2010} by introducing near duplicates. In this study the notion of documentation phrase is used to denote the near duplicate. Parametrization is used to define variative parts of duplicates, similarly to our approach ~\cite{koznov2008,romanovsky2008}. However, the authors left the problem of near duplicate detection untouched. 

In~\cite{nosal2016} Nos\'{a}l and Porub\"{a}n present the results of a case study in which they searched for exact duplicates in internal documentation (source code comments) of an open source project set. They used a modified copy/paste detection tool, which was originally developed for code analysis and found considerable number of text duplicates. However, near duplicates were not considered in this paper.

Wingkvist et al. adapted a clone detection tool to measure the document uniqueness in a collection~\cite{wingkvist2010}. The authors used found duplicates for documentation quality estimation. However, they did not address near duplicate detection. 

The work of Juergens et al.~\cite{juergens2010} is the closest one to our research and presents a case study for analyzing redundancy in requirement specifications. The authors analyze 28 industrial documents. At the first step, they found duplicates using a clone detection tool. Then, the authors filtered the found duplicates by manually removing false positives and performed a classification of the results. They report that the average duplicate coverage of documents they analyzed is 13.6\%: some documents have a low coverage (0.9\%, 0.7\% and even 0\%), but there are ones that have a high coverage (35\%, 51.1\%, 71.6\%). Next, the authors discuss how to use discovered duplicates and how to detect related duplicates in the source code. The impact of duplicates on the document reading process is also studied. Furthermore, the authors propose a classification of meaningful duplicates and false positive duplicates. However, it should be noted that they consider only requirement specifications and ignore other kinds of software documentation. Also, they do not use near duplicates.

Rago et al.~\cite{rago2016} apply natural language processing and machine learning techniques to the problem of searching duplicate functionality in requirement specifications. These documents are considered as a set of textual use cases; the approach extracts sequence chain (usage scenarios) for every use case and compares the pairs of chains to find duplicate subchains. The authors evaluate their approach using several industrial requirement specifications. It should be noted that this study considers a very special type of requirement specifications, which are not widely used in industry. Near duplicates are also not considered. 

Algorithms for duplicate detection in textual content have been developed in several other areas. Firstly, the information retrieval community considered several near-duplicate detection problems: document similarity search \cite{Huang:2013,Williams:2013}, (local) text reuse detection \cite{Zhang:2010,AbdelHamid:2009}, template detection in web document collections \cite{Ramaswamy:2004,Gibson:2005}. Secondly, the plagiarism detection community also extensively studied the detection of similarity between documents \cite{Valles:2011,Barron-Cedeno:2013}. Thus, there are a number of efficient solutions. However, the majority of them are focused on document-level similarity, with the goal of attaining high performance on the large collections of documents. More importantly, these studies differ from ours by ignoring duplicate meaningfulness. Nonetheless, adapting these methods for the near-duplicate search in software documentation could improve the quality metrics of the search. It is a promising avenue for further studies.

The theoretical computer science community also addressed the duplicate detection problem. However, the primary focus of this community was the development of (an approximate) string matching algorithms. For example, in order to match two text fragments of an equal length, the Hamming distance was used~\cite{smyth2003}. The Levenshtein distance~\cite{levenshtein1965} is employed to account not only for symbol modification, but also for symbol insertion and removal. For this metric it is possible to handle all the three editing operations at the cost of performance~\cite{wagner1974}. Significant performance optimizations to string pattern matching using the Levenshtein distance were applied in the fuzzy Bitap algorithm~\cite{manber.wu1992} which was later optimized to handle longer patterns efficiently~\cite{myers1999}. Similarity preserving signatures like MinHash~\cite{broder1997} can be used to speed up pattern matching for approximate matching problem. Another indexing approach is proposed in~\cite{sigmod16}, with an algorithm for the fast detection of documents that share a common sliding window with the query pattern but differ by at most $\tau$ tokens. While being useful for our goals in general, these studies do not relate to this paper directly. All of these studies are focused on the performance improvement of simple string matching tasks. Achieving high performance is undoubtedly an important aspect, but these algorithms become less necessary when employed for documents of 3--4 MBs of plain text --- a common size of a large industrial documentation file.

Various techniques have been employed to detect near duplicate clones in a source code. SourcererCC~\cite{sajnani2016} detects near duplicates of code blocks using a static bag-of-tokens strategy that is resilient to minor differences between code blocks. Clone candidates of a code block are queried from a partial inverted index for better scalability. DECKARD~\cite{jiang2007} computes certain characteristic vectors of code to approximate the structure of Abstract Syntax Trees in Euclidean space. Locality sensitive hashing~(LSH)~\cite{Indyk:1998:ANN:276698.276876} is used to group similar vectors with the Euclidean distance metric, forming clones. NICAD~\cite{cordy2011} is a text-based near duplicate detection tool that also uses tree-based structural analysis with a lightweight parsing of the source code to implement flexible pretty-printing, code normalization, source transformation and code filtering for better results. However, these techniques are not directly capable of detecting duplicates  in text documents as they involve some degree of parsing of the underlying source code for duplicate detection. Suitable customization for this purpose can be explored in the future.


\section{Background}\label{sec:background}
\subsection{Exact duplicate detection and Clone Miner}
Not only documentation, but also software itself is often developed with a lot of copy/pasted information. To cope with duplicates in the source code, software clone detection methods are used. This area is quite mature; a systematic review of clone detection methods and tools can be found in~\cite{rattan2013}.
In this paper, the Clone Miner~\cite{basit2009} software clone detection tool is
used to detect exact duplicates in software documentation. Clone Miner is a token-based source code clone detector. A token in the context of text documents is a single word separated from other words by some separator: ‘.’, ‘(’, ‘)’, etc. For example, the following text fragment consists of 2 tokens: ``FM registers''. Clone Miner considers input text as an ordered collection of lexical tokens and applies suffix array-based string matching algorithms~\cite{abouelhoda2004} to retrieve the repeated parts (clones).
In this study we use the Clone Miner tool. We have selected it for its simplicity and its ability to be easily integrated with other tools using a command line interface.
\subsection{Basic near duplicate detection and refactoring}\label{sec:basic}

We have already demonstrated in \textbf{Section 1} that near duplicate detection in software documentation is an important research avenue. In our previous studies~\cite{psi15,prog16}, we presented an approach that offers a partial solution for this problem. At first, similarly to Juergens et al.~\cite{juergens2010}, Wingkvist et al.~\cite{wingkvist2010}, we applied software clone detection techniques to exact duplicate detection \cite{kio2012}. Then, in~\cite{psi15,prog16} we proposed an approach to near duplicate detection. It is essentially as follows: having exact duplicates found by Clone Miner, we extract sets of duplicate groups where clones are located close to each other. For example, suppose that the following phrase can be found in the text 5 times with different variations (various port numbers): ``inet daemon can listen on ... port and then transfer the connection to appropriate handler''. In this case we have two duplicate groups with 5 clones in each group: one group includes the text ``inet daemon can listen on'', while the other includes ``port and then transfer the connection to appropriate handler''. We combine these duplicate groups into a group of near duplicates: every member of this group has one variation to capture different port numbers. In these studies we developed this approach only for one variation, i.e. we ``glued'' only pairs of exact duplicate groups.

\begin{figure}
\centering
\includegraphics[scale=0.4]{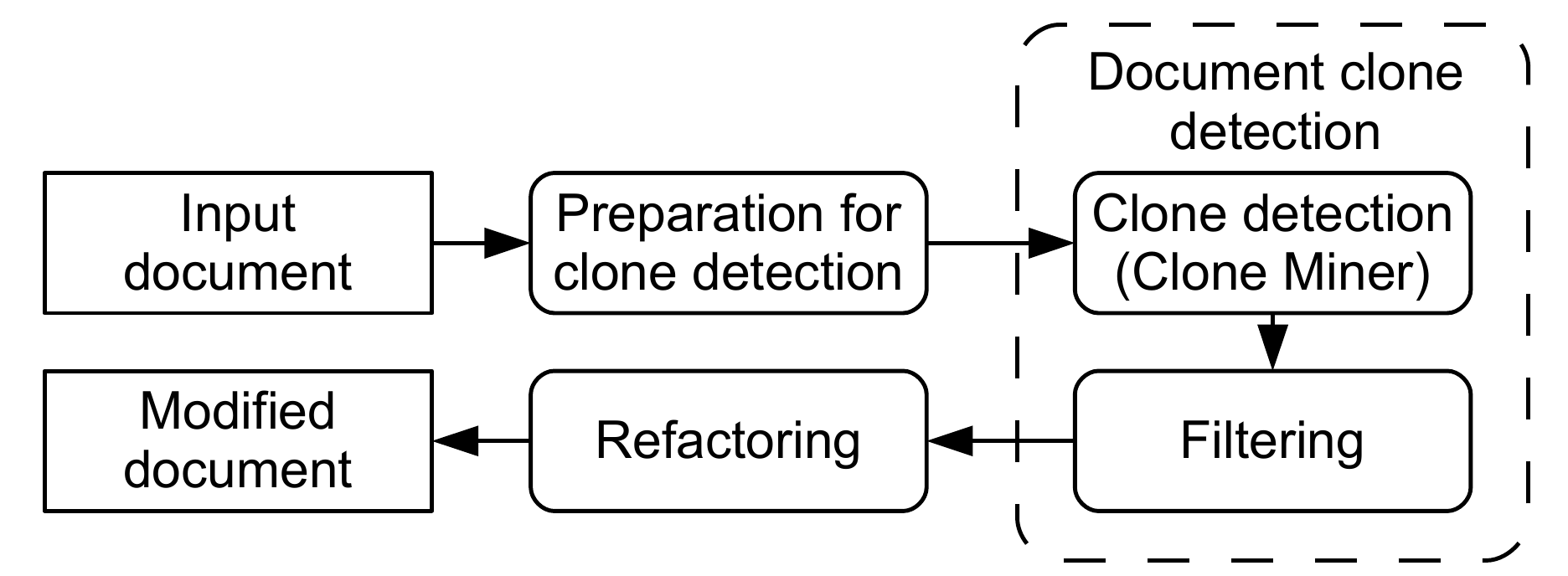}
\caption{The process of near duplicate search}
\label{fig:process2015}
\end{figure}

Then it is possible to apply the adaptive reuse technique~\cite{bassett1997, jarzabek2003} to near duplicates and to perform automatic refactoring of documentation. Our toolkit~\cite{DRT} allows to create reusable text fragment definition templates for the selected near duplicate group. Then it is possible to substitute all the occurrences of the group’s members with parameterized references to definition. The overall scheme of the process is shown in Fig.~\ref{fig:process2015}.

The experiments with the described approach showed a considerable number of interesting near duplicates. Thus, we decided to generalize the algorithm from~\cite{psi15,prog16}
in order to allow an arbitrary number of exact near-duplicates combinations. This will allow to have several variation parts in the resulting near duplicates. The generalized algorithm is described below.

\section{Near duplicate detection algorithm}\label{sec:alg}
\subsection{Definitions}\label{sec:vargroups} 
Let us define the terms necessary for describing the proposed algorithm. We consider document $D$ as a sequence of symbols. Any symbol of $D$ has a coordinate corresponding to its offset from the beginning of the document, and this coordinate is a number belonging to $[1, length(D)]$ interval, where $length(D)$ is the number of symbols in $D$.

\begin{definition}
For $D$ we define a \textbf{text fragment} as an occurrence of some text substring in $D$.
Hence, each text fragment has a corresponding integer interval $[b, e]$, where $b$ is the coordinate of its first symbol and $e$ is the coordinate of its last symbol. For text fragment $g$ of document $D$, we say that $g \in D$.
\end{definition}

Let us introduce the following sets: $D^*$ is a set of all text fragments of $D$, $I^D$ is a set of all integer intervals within interval $[1, length(D)]$,  $S^D$ is a set of all strings of $D$.

Also, let us introduce the following notations:
\begin{itemize}\setlength\itemsep{-0.2em}
\item  $[g]: D^* \rightarrow I^D$ is a function that takes text fragment $g$ and returns its interval.
\item  $str(g): D^* \rightarrow S^D$ is a function that takes text fragment $g$ and returns its text.
\item $\overline{I}:  I^D \rightarrow D^*$ is a function that takes interval $I$ and returns corresponding text fragment.
\item $|[b, e]|: I^D \rightarrow [{0, length(D)}]$ is a function that takes interval $[g]= [b, e]$ and returns its length as $\left|[g]\right|=e-b+1$. For simplicity, we will use $|g|$ notion instead of $|[g]|$.
\item For any $g^1, g^2 \in D$ we consider their intersection $g^1 \cap g^2$ as intersection of corresponding intervals $[g^1] \cap [g^2]$, and $g^1 \subset g^2$ implies $[g^1] \subset [g^2]$.
\item We define the binary predicate $Before$ on $D^* \times D^*$, which is true for text fragments $g^1, g^2 \in D$, iff $e^1<b^2$, when $[g^1 ]=[b^1, e^1 ], [g^2]=[b^2, e^2]$.
\end{itemize}

\theoremstyle{definition}
\begin{definition}
Let us consider a set $G$ of text fragments of $D$ such that $\forall g^1, g^2 \in G$ $(str(g^1) = str(g^2)) \wedge (g^1 \cap g^2 = \emptyset)$. We name those fragments as \textbf{exact duplicates} and $G$ as \textbf{exact duplicate group} or \textbf{exact group}. We also denote number of elements in $G$ as $\#G$.
\end{definition}

\begin{definition}\label{def:variation group}
For ordered set of exact duplicate groups $G_1,\ldots, G_N$, we say that it forms \textbf{variational group} $\langle G_1,\ldots, G_N \rangle$ when the following conditions are satisfied:
\begin{enumerate}
\item $\#G_1 =\ldots = \#G_N$. 
\item\label{item:before} Text fragments having similar positions in different groups, occur in the same order in document text:
$\forall g_i^{k} \in G_i$  $\forall g_j^{k} \in G_j$
$((i < j) \Leftrightarrow Before(g_i^{k}, g_j^{k}))$, and
\newline $\forall~k~\in~\{1,~\ldots,~N-1\} Before (g_N^{k}, g_1^{k+1})$.
\end{enumerate}
We also say that for any $G_k$ of this set $G_k \in VG$.
\end{definition}

\begin{remark}
According to condition~\ref{item:before} of definition~\ref{def:variation group}, $\forall g_i^k \in G_i, \forall g_j^k \in G_j (i \neq j \Rightarrow g_i^k \cap g_j^k = \emptyset)$.
\end{remark}

\begin{remark}\label{rem:variation group}
When $VG= \langle G_1,\ldots, G_N \rangle$ and $VG'= \langle G'_1,\ldots, G'_M \rangle$, are variational groups, $\langle VG, VG' \rangle =\langle G_1,\ldots, G_N, G'_1,\ldots, G'_M \rangle$ is also a variational group in case when is satisfies definition~\ref{def:variation group}.
\end{remark}

For example, suppose that we have $VG = \langle G_1, G_2, G_3 \rangle$ and
each of $G_i$ consists of three clones
$g_i^k,\ i \in \{1, 2, 3\}$. Then, these clones appear in the text in the following order: $
g_1^1 \ldots g_2^1 \ldots g_3^1
\ldots\ldots\allowbreak
g_1^2 \ldots g_2^2 \ldots g_3^2
\ldots\ldots\allowbreak
g_1^3 \ldots g_2^3 \ldots g_3^3 $.
Next, it should be possible to compute the distance between variations or exact duplicate groups. This is required to support group merging inside our algorithm which selects several closest groups to form a new one. Thus, a distance function should be defined.

\begin{definition} \textbf{Distance between text fragments} for any $g^1, g^2 \in D$ is defined as follows:
\begin{equation}
dist(g^1, g^2) = \begin{cases}
0, & ~ g^1 \cap g^2 \ne \emptyset,\\
b^2 - e^1+1, & ~ Before(g^1, g^2),\\
b^1 - e^2+1, & ~ Before(g^2, g^1),\\
\end{cases}
\end{equation} 
where $[g^1] = [b^1, e^1]$ and $[g^2]=[b^2, e^2]$.
\end{definition}

\begin{definition}\label{def:gdist}
\textbf{Distance between exact groups} $G_1$ and $G_2$, having $\#G_1 = \#G_2$, is defined as follows:
\begin{equation}
dist(G_1, G_2) = \max\limits_{k\in\{1,\ldots,\#G_1\}}dist(g_1^k, g_2^k) 
\end{equation}
\end{definition}

\begin{definition}\label{def:vgdist}
\textbf{Distance between variational groups} $VG_1$ and $VG_2$, when there are $G_1 \in VG_1, G_2 \in VG_2: \#G_1=\#G_2$, is defined as follows:
\begin{equation}
dist(VG_1, VG_2) =\\\max\limits_{G_1 \in VG_1, G_2 \in VG_2}dist(G_1,G_2)
\end{equation}
\end{definition}

\begin{definition}\label{def:gl}
\textbf{Length of exact group} $G$ is defined as follows:
 $length(G) = \sum\limits_{k=1}^{\#G} (e^k - b^k+1)$, where $g^k \in G, [g^k] = [b^k, e^k]$.
\end{definition}

\begin{definition}\label{def:vgl}
\textbf{Length of variational group} $VG = \langle G_1,\ldots, G_N \rangle$ is defined as follows:
\begin{equation}
length(VG) = \sum\limits_{i=1}^{N}length(G_i)\end{equation}
\end{definition}

\begin{definition}\label{def:ndg}
\textbf{Near duplicate group} is such a variational group $\langle G_1,\ldots, G_N\rangle$ that satisfies following condition for $\forall k \in \{1,\ldots,\#G_1\}$:
\begin{equation}\label{eq:bassett15}
\begin{split}
\sum\limits_{i=1}^{N-1}dist(g_i^k,g_{i+1}^k) \le 0.15 * \sum\limits_{i=1}^{N}|g_i^k|.
\end{split}
\end{equation}
\end{definition}

 This definition is constructed according to the near duplicate concept from~\cite{bassett1997}: variational part of near duplicates with similar information (\emph{delta}) should not exceed 15\% of their exact duplicate (\emph{archetype}) part.


\begin{remark}\label{rem:single}
An exact group $G$ can be considered as a variational one formed by itself: $\langle G \rangle$.
\end{remark}

\begin{definition} 
Consider near duplicate group $\langle G_1, G_2\rangle$, where $G_1$ and $G_2$ are exact groups. We assume that this group contains a single \textbf{extension point}, and the text fragments contained in positions $[e_1^k + 1, b_2^k - 1]$ are called \textbf{extension point values}. In the general case, a near duplicate group $\langle G_1,\ldots, G_N\rangle$ has $N-1$
\emph{extension points}.
\end{definition}

\begin{definition}\label{def:nearbyndg}
Consider two near duplicate groups $G=\langle G_1,\ldots, G_n \rangle$ and $G'=\langle G'_1,\ldots G'_m \rangle$. Suppose that they form a variational group $\langle G_1,\ldots, G_n, G'_1,\ldots, G'_m \rangle$ or $\langle G_1,\ldots, G_n, G'_1,\ldots, G'_m \rangle$, which in turn is also a near duplicate group. In this case, we call $G$ and $G'$ \textbf{nearby groups}.
\end{definition}

\begin{definition}\label{def:nearbynd}
\textbf{Nearby duplicates} are duplicates belonging to nearby groups.
\end{definition}

\begin{remark}
Due to remark~\ref{rem:single}, definition \ref{def:nearbynd} is applicable to both near and exact duplicates.
\end{remark}


\subsection{Algorithm description}

The algorithm that constructs the set of near duplicate groups ($SetVG$) is presented below. Its input is the set of exact duplicate groups ($SetG$) belonging to document $D$. It employs an interval tree~--- a data structure whose purpose is to quickly locate intervals that intersect with a given interval. Initially, the $SetG$ set is created using the Clone Miner tool. The core idea of our algorithm is to repeatedly find and merge nearby exact groups from $SetG$. At each step, the resulting near duplicate groups are added to $SetVG$. Let us consider this algorithm in detail.

\en
\renewcommand{\algorithmcfname}{\iflanguage{russian}{Алгоритм}{Algorithm}}

\SetInd{0.125em}{0.75em}
\begin{algorithm}[h]\label{alg:main}
\caption{\iflanguage{russian}{Конструирование групп неточных повторов}{Near Duplicate Groups Construction}}
\LinesNumbered
\DontPrintSemicolon
\SetKwInput{KwIn}{\iflanguage{russian}{Входные данные}{Input data}}
\SetKwInput{KwOut}{\iflanguage{russian}{Результат}{Result}}
\KwIn{$SetG$} 
\KwOut{$SetVG$}
$SetVG \gets \emptyset$

$Initiate()$

\Repeat { $SetNew = \emptyset$ } {
	$SetNew \gets  \emptyset$

	\ForEach{$G \in  SetG \cup SetVG$}{

		$SetCand \gets NearBy(G$)

		\If{ $SetCand \ne \emptyset$ }{
			$G' \gets GetClosest(G, SetCand)$
			
			$Remove (G,G')$
			
			\uIf{$Before(G,G')$ }{
				$SetNew \gets SetNew \cup \{\langle G, G' \rangle\}$
			}
			\Else{
				$SetNew \gets SetNew \cup \{\langle G', G \rangle\}$
			}
		}
	}
	
	$Join (SetVG, SetNew)$
}

$SetVG \gets SetVG \cup SetG$
\en
\end{algorithm}

The initial interval tree for $SetG$ is constructed using the $Initiate()$ function (line 2). The core part of the algorithm is a loop in which new near duplicate groups are constructed (lines 3--18).
This loop repeats until we can construct at least one near duplicate group, i.e. the set of newly constructed near duplicate groups ($SetNew$) is not empty (line 18).
Inside of this loop, the algorithm cycles through all groups of $SetG \cup SetVG$. For each of them, the $NearBy$ function returns the set of nearby groups $SetCand$ (lines 5, 6), which is then used for constructing near duplicate groups. 
Later, we will discuss this function in more detail and prove its correctness, i.e. that it actually returns groups that are close to $G$. 
Next, the closest group to $G$, denoted $G'$, is selected from $SetCand$ (line 8) and a variational group $\langle G, G' \rangle$ or $\langle G', G \rangle$ is created. This group is added into $SetNew$ (lines 10--14). Since $G$ and $G'$ are merged and therefore cease to exist as independent entities, they are deleted from $SetG$ and $SetVG$ by the $Remove$ function (line 9). 
Next, the $Join$ function adds $SetNew$ to $SetVG$ (line 17). It is essential to note that the $Remove$ and $Join$ functions perform some auxiliary actions described below.

In the end of the algorithm $SetG$ is added to $SetVG$. The result ~--- $SetVG$~--- is presented as the algorithm's output. This step is required in order for the output to contain not only near duplicate groups, but also exact duplicate groups which have not been used for creation of near duplicate ones (line 19).

Let us describe the functions employed in this algorithm.

The $Initiate()$ function builds the interval tree. The idea of this data structure is the following.

Suppose we have $n$ natural number intervals, where $b_1$ is the minimum and $e_n$ is the maximum value of all interval endpoints, and $m$ is the midpoint of $[b_1, e_n]$.
The intervals are divided into three groups: fully located to the left of $m$, fully located to the right of $m$, and intervals containing $m$. The current node of the interval tree stores the last interval group and references to its left and right child nodes containing the intervals to the left and to the right of $m$ respectively.
This procedure is repeated for each child node. Further details regarding the construction of an interval tree can be found in ~\cite{deberg2000,preparata1985}.

In this study, we build our interval tree from the extended intervals that correspond to the exact duplicates found by CloneMiner. 
 These extended  intervals are obtained as follows: original intervals belonging to exact duplicates are enlarged by 15\%. For example, if $[b, e]$ is the initial interval, then an extended one is $[b - 0.15 * (e-b+1), e + 0.15 * (e-b+1)]$. We will denote the extended interval that corresponds to the exact duplicate $g$ as $\updownarrow g$. We also modify our interval tree as follows: each stored interval keeps the reference to the corresponding exact duplicate group.

The $Remove$ function removes groups from sets and their intervals from the interval tree. The interval deletion algorithm is described in references~\cite{deberg2000, preparata1985}.

The $Join$ function, in addition to the operations described above, adds intervals of the newly created near duplicate group $G =\langle G_1,\ldots, G_N \rangle$ to the interval tree. The standard insertion algorithm described in references~\cite{deberg2000,preparata1985} is used. Extended intervals added to the tree of each near duplicate $g^k =(g_1^k,..., g_N^k)$, where $k\in\{1,\ldots,\#G_1\}$, have the form of $[b_1^k - x^k, e_N^k + x^k]$, where $x^k = 0.15 * \sum\limits_{i=1}^N |g_i^k| - \sum\limits_{i=1}^{N-1} dist(g_i^k, g_{i+1}^k)$.
We will denote this extended interval of $g^k$ (now, a near duplicate) as ${\updownarrow g^k}$ as well.

The $NearBy$ function selects nearby groups for some group $G$ (its parameter). To do this, for each text fragment from $G$ a collection of intervals that intersect with its interval is extracted.\emph{Text fragments that correspond to these intervals turn out to be neighboring to the initial fragment}, i.e. for them, condition~(\ref{eq:bassett15}) is satisfied.
The retrieval is done using the interval tree search algorithm~\cite{deberg2000,preparata1985}. We construct the $GL_1$ set, which contains groups that are expected to be nearby to $G$:

\vspace*{-2mm}
\begin{equation}
\begin{split}
GL_1 (G) = \{G'| (G' \in SetG \cup SetVG) \wedge\\
\exists g \in G,\,g' \in G':\;\updownarrow g  \cap \updownarrow g' \ne \emptyset.
\end{split}
\end{equation}

That is, the $GL_1$ set consists of groups that contain at least one duplicate that is close to at least one duplicate from $G$. Then, only the groups that can form a variational group with $G$ are selected and placed into the $GL_2$ set:

\vspace*{-5mm}
\begin{equation}
\begin{split}
GL_2 (G) = \{G' | G' \in GL_1 \wedge
(\langle G, G' \rangle \mbox{or}
\\
\langle G', G \rangle \mbox{is variational group})\}.
\end{split}
\end{equation}

Finally, the $GL_3$ set (the $NearBy$ function's output) is created. The only groups placed in this set are those from $GL_2$ whose all elements are close to corresponding elements of $G$:

\vspace*{-5mm}
\begin{equation}
\begin{split}
GL_3 (G) = \{G'|G'\in GL_2\,\wedge\\
\forall k\in\{1,\ldots,\#G\}:\;\updownarrow g^k \cap \updownarrow g^{\prime k} \ne\emptyset\}
\end{split}
\end{equation}

\begin{theorem}\label{th:tnb} Suggested algorithm detects near duplicate groups that conform to definition ~\ref{def:ndg}.
\end{theorem}

It is easy to show by construction of $NearBy$ that for some group $G$ it returns the set of its nearby groups (see definition~\ref{def:nearbyndg}). That is, each of these groups can be used to form a near duplicate group with $G$. Then for the set the algorithm selects the group closest to $G$ and constructs a new near duplicate group. The correctness of all intermediate sets and other used functions is immediate from their construction methods.

\en

\section{Evaluation}\label{sec:eval}
The proposed algorithm was implemented in the Duplicate Finder Toolkit \cite{DRT}. Our prototype uses the \texttt{intervaltree} library~\cite{pyintervaltree} as an implementation of the interval tree data structure. 

We have evaluated 19 industrial documents belonging to various types: requirement specification, programming guides, API documentation, user manuals, etc. (see Table 1). The size of the evaluated documents is up to 3 Mb. 

Our evaluation produced the following results. The majority of the duplicate groups detected are exact duplicates (88.3--96.5\%). Groups having one variation point  amount to 3.3--12.5\%, two variation points – to 0--1.7\%, and three variation points – to less 1\%, etc. A few near duplicates with 11, 12 13, and 16 variation points also were detected. 
We performed a manual analysis of the automatically detected near duplicates for Linux Kernel Documentation (programming guide, document 1 in the Table~\ref{tab:eval})~\cite{linuxkd2013}. We found 70 meaningful text groups (5.4\%), 30 meaningful groups for the code example (2.3\%), and 1191 false positive groups (92.3\%). We found 21 near duplicate groups, i.e. 21\% of the meaningful duplicate groups. Therefore, the share of near duplicates significantly increases after discarding false positives. 

\begin{table*}[!t]
\centering
\caption{Near-duplicate groups detected}
\label{tab:eval}
\includegraphics[scale=0.8]{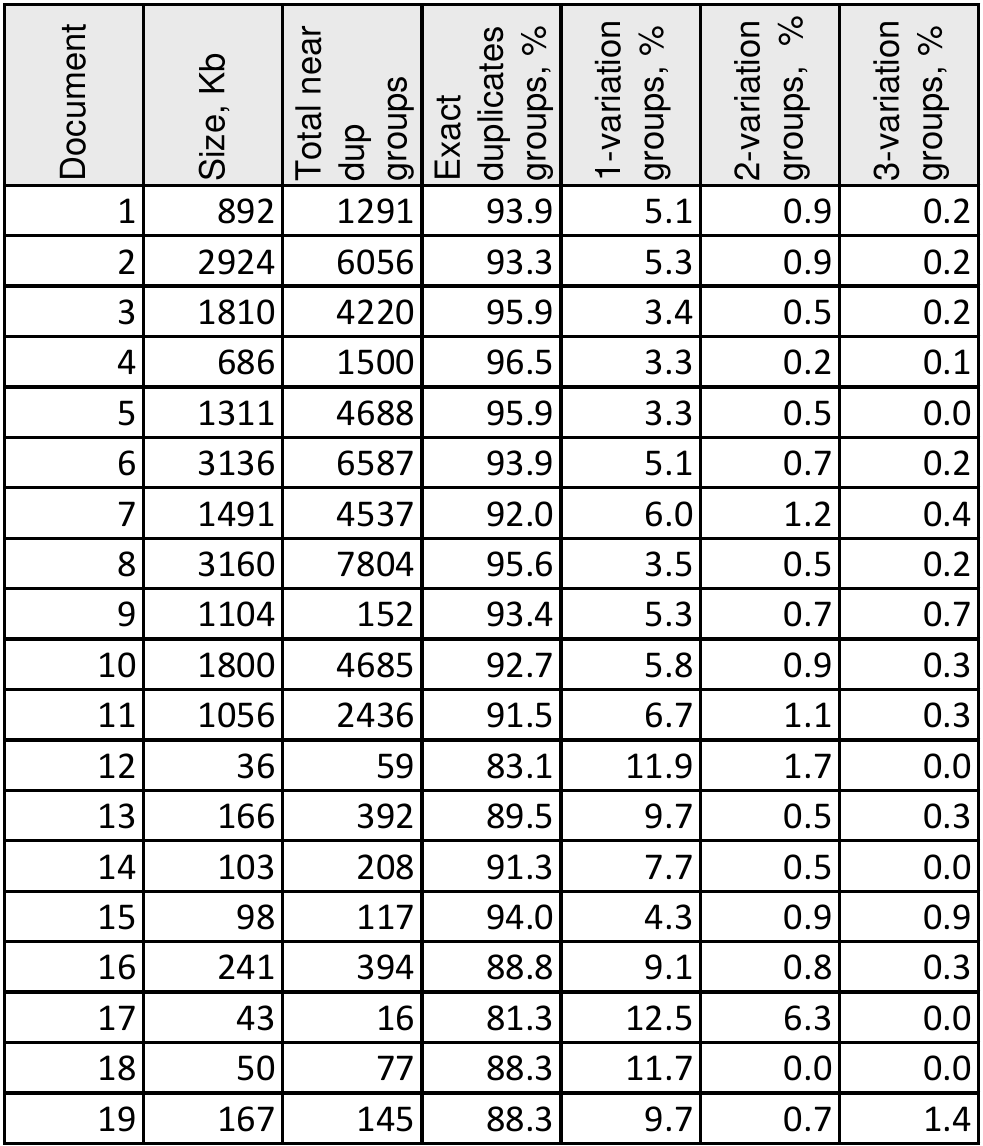}
\end{table*}

Having analyzed the evaluation results, we can make the following conclusions:
\begin{enumerate}\ru

\item During our experiments we did not manage to find any near duplicates in considered documents that were not detected by our algorithm. However, we should note that the claim of the algorithm's high recall needs a more detailed justification.

\item Analyzing the Linux Kernel Documentation, we have concluded that it does not have any cohesive style: it was created sporadically by different authors. Virtually all its duplicates are situated locally, i.e. close to each other. For example, some author created a description of some driver's functionality using copy/paste for its similar features. At the same time, another driver was described by a different author who did not use the first driver's description at all. Consequently, there are practically no duplicates that are found throughout the whole text. Examples, warnings, notes, and other documentation elements that are preceded by different introductory sentences are not styled cohesively as well. Thus, our algorithm can be used for analyzing the degree of documentation uniformity.

\item The algorithm performs well on two-element groups, finding near duplicate groups with a different number of extension points. It appears that, in general, there are way fewer near duplicate groups with more than two elements.

\item Many detected duplicate groups consist of figure and table captions, page headers, parts of the table of contents and so on~--- that is, they are not of any interest to us. Also, many found duplicates are scattered across different elements of document structure, for example, a duplicate can be a part of a header and a small fragment of text right after it. These kinds of duplicates are not desired since they are not very useful for document writers. However, they are detected because currently document structure is not taken into account during the search.

\item The 0.15 value used in detecting near duplicate groups does not allow to find some significant groups (mainly small ones, 10--20 tokens in size). It is possible that it would be more effective to use some function instead of a constant, which could depend, for example, on the length of the near duplicate.

\item Moreover, often the detected duplicate does not contain variational information that is situated either in its end or in its beginning. Sometimes it could be beneficial to include it in order to ensure semantic completeness. To solve this problem, a clarification and a formal definition of semantic completeness of a text fragment is required. Our experiments show that this can be done in various ways (the simplest one is ensuring sentence-level granularity, i.e. including all text until the start/end of sentence).

\end{enumerate}

\section{Conclusion}
In this paper the formal definition of near duplicates in software documentation is given and the algorithm for near duplicate detection is presented. An evaluation of the algorithm using a large number of both commercial and open source documents is performed. 

The evaluation shows that various types of software documentation contain a significant number of exact and near duplicates. A near duplicate detection tool could improve the quality of documentation, while a duplicate management technique would simplify documentation maintenance. 

Although the proposed algorithm provides ample evidence on text duplicates  in industrial documents, it still requires improvements before it can be applied to real-life tasks. The main issues to be resolved are the quality of the near duplicates detected and a large number of false positives. Also, a detailed analysis of near duplicate types in various sorts of  software documents should be performed. 

\bibliographystyle{prog_en}
\bibliography{references}
\end{document}